\documentclass[useAMS, usenatbib]{mn2e}

\usepackage{psfig}

\usepackage{macros_desai}
\begin{document}


\title[The  Cluster Galaxy  Circular Velocity  Function]  {The Cluster
Galaxy Circular Velocity Function}

\author[V.     Desai    \etal]    {V.~Desai    \thanks{E-mail:    {\tt
desai@astro.washington.edu}},  J.~J.~Dalcanton \thanks{Alfred P.~Sloan
Research   Fellow},    L.~Mayer,   D.~Reed,   T.~Quinn,   F.~Governato
\thanks{David E.~Brooks  Research Fellow}\\ University  of Washington Department of Astronomy, Box 351580, Seattle WA 98195-1580, USA}

\date{}             \pagerange{\pageref{firstpage}--\pageref{lastpage}}
\pubyear{2003}

\maketitle

\label{firstpage}


\begin{abstract}
We present galaxy circular velocity functions (GCVFs) for 34 low
redshift ($z \la 0.15$) clusters identified in the Sloan Digital Sky
Survey (SDSS), for fifteen clusters drawn from dark matter simulations
of hierarchical structure growth in a $\Lambda$CDM cosmology, and for
$\sim$29,000 SDSS field galaxies.  We find that the observed and
simulated cluster GCVFs take the form of a power law.  The cumulative
GCVFs of the simulated clusters are very similar across a wide range
of cluster masses, provided individual subhalo circular velocities are
scaled by the circular velocities of the parent cluster.  Once all
sources of error are accounted for, the intrinsic scatter in the
cumulative, scaled observed cluster GCVF is consistent with the
simulations.  The slope of the observed cluster GCVF is $\sim$-2.5,
independent of cluster velocity dispersion.  The average slope of the
simulated GCVFs is somewhat steeper, although formally consistent
given the errors.  Using our highest resolution cluster, we find that
the effects of baryons on galaxy rotation curves is to flatten the
simulated cluster GCVF into better agreement with observations.
Finally, we find that the field GCVF deviates significantly from a
power law, being flatter than the cluster GCVF at circular velocities
less than 200 $\kms$, and steeper at circular velocities greater than
200 $\kms$.

\end{abstract}


\begin{keywords}
cosmology: theory --- cosmology:  observations --- galaxies: halos ---
galaxies:  clusters ---  methods:  {\it N}-body  simulations ---  dark
matter.
\end{keywords}


\section{Introduction}
\label{sec:intro}

\indent In  the widely  accepted  hierarchical clustering  model of  structure
formation, adiabatic  fluctuations in  the initial density  field grow
and  merge  into  increasingly   larger  dark  matter  halos.   Simple
inflationary approaches model the initial power spectrum of fluctuations
as Gaussian and  scale-invariant, while cosmological parameters and
the nature of  the dark matter control its  subsequent evolution.  The
resulting mass  function of dark  matter halos can be  predicted using
{\it  N}-body   simulations  \citep{Sheth99,  Bode01,   Jenkins01, Reed03}  or
analytic  methods  based  upon Press-Schecter  theory  \citep{Press74,
Bond91,  Bower91, Lacey93,  Sheth01}.  The  observed mass  function is
therefore  an important  constraint on  fundamental properties  of the
universe.  In the  following, we focus on the  mass function on galaxy
scales. 

Comparing predicted mass functions to observations requires mapping
the observable properties of galaxies onto the masses of their dark
matter halos.  In one technique, semi-analytic models (SAMs) are used
to predict the luminosities of galaxies within halos of different
masses, by parametrizing models of the complicated processes of gas cooling,
star formation, stellar evolution, and feedback.  These techniques
efficiently explore a wide variety of models.  However, their utility
is limited by the extent to which the simple recipes capture the
relevant physics.

An alternative to SAMs utilizes the empirical relationship
between velocities and luminosities in both spiral and elliptical
galaxies to transform the observed galaxy luminosity function into a
galaxy circular velocity function (GCVF), which is closely related to
the mass function \citep{Cole89, Shimasaku93, Gonzalez00, Kochanek01}.
While this approach avoids a number of assumptions required to
assign luminosities to dark matter halos, baryons cannot be completely
ignored.  The baryonic mass distribution affects a galaxy's circular
velocity profile significantly, while also changing the equilibrium
configuration of the host dark matter halo through adiabatic
contraction \citep{Blumenthal86}.  Baryons also determine the
probability that a galaxy will be included in a given observational
sample.  However, the circular velocity directly traces the total
enclosed mass of a galaxy, which is dominated by dark matter at
sufficiently large radii.  Therefore, models of the GCVF are less
sensitive to inaccuracies in our understanding of star formation,
stellar evolution, and feedback than those of the luminosity function.
The observed GCVF therefore better constrains the mass function.

Recent comparisons between observed and theoretical field GCVFs
indicate that low-mass halos with circular velocities $\la$200$\kms$
are over-predicted by simulations \citep{Gonzalez00, Kochanek01}.  It
is now possible to extend the analysis of the GCVF to galaxies within
clusters.  The current generation of {\it N}-body simulations
\citep{Ghigna98, Tormen98, Klypin99clus, Okamoto00, Springel01} has
achieved the force and mass resolution necessary to capture the
effects of tidal stripping, dynamical friction, and galaxy harassment,
essential for predicting the GCVF in dense environments.
\citet{Moore99} used the Tully-Fisher relation to determine circular
velocities for galaxies in the Virgo Cluster.  They compared the
resulting distribution to that in a standard CDM simulation of a
Virgo-like cluster, and found good agreement down to $\sim$50 $\kms$
without accounting for the effects of baryons on galaxy rotation
curves.  Coupling dark matter merging trees with SAMs
which include strong feedback, \citet{Springel01} were able to
reproduce a wide variety of cluster properties.  Although they did not
directly compute the cluster GCVF, the agreement between a variety of
observed and modeled cluster relations (morphology-density, luminosity
function, Tully-Fisher, Faber Jackson) suggests that they would have
found agreement for the cluster GCVF as well.  Although these works
represent significant progress in testing simulations of cluster
substructure against observations, larger samples of both simulations
and observations are required to assess whether the results are typical,
and over what cluster mass range.

Toward this goal, we determine the GCVFs for 34 nearby clusters from
the Sloan Digital Sky Survey (SDSS).  To compare the observed clusters
to theoretical predictions, we assemble a sample of 15 high resolution
cluster simulations, from which we measure the theoretical cluster
GCVF.  We also repeated the GCVF analysis for $\sim$29,000 galaxies
from the SDSS, to allow a comparison to the field.

Our method for determining the GCVFs for the SDSS sample is described
in \S{\ref{sec:obs}}.  In \S{\ref{sec:sims}} we describe the
simulations.  Our results are discussed in \S{\ref{sec:results}}.
First we discuss the shape of the cluster GCVF as related to that of
the cluster luminosity function (\S{\ref{sec:shape}}).  Second, we
investigate the claim of \citet{Moore99} that the velocity
distribution of substructure on all mass scales is self-similar when
scaled by the host halo velocity (\S{\ref{sec:selfsim}}).  Third, we
present power-law fits to the cluster GCVF, quantify its dependence on
cluster velocity dispersion, and make a comparison to dark matter
cluster simulations (\S{\ref{sec:trends}}).  Fourth, we examine how
the adiabatic contraction of dark matter halos due to baryons affects
the comparison between observed and simulated GCVFs
(\S{\ref{sec:hires}}).  Finally, we compare the observed cluster GCVF
to that measured for the field (\S{\ref{sec:field}}).  A summary of
our results can be found in \S{\ref{sec:summary}}.

Throughout we adopt $H_{{\rm 0}} = 70\kms {\rm Mpc}^{-1}$,
$\Lambda = 0.7$, and $\Omega = 0.3$.

\section{Methods:  Construction of the Observed Cluster Galaxy Circular Velocity Function}
\label{sec:obs}

To  construct  the observed  cluster  GCVF,  we  use existing  cluster
catalogs and data from the SDSS.  For each
cluster,  we  calculate  the  characteristic circular  velocities  for
galaxies  within  the  virial  radius,  using  the  Tully-Fisher  and
Fundamental Plane relations.  Note that this is done galaxy-by-galaxy,
rather than by transforming the  luminosity function, as has been done
previously  \citep{Cole89,  Gonzalez00,  Kochanek01}.  In
this  section,  we  describe  the  SDSS,  the  cluster  catalogs,  our
determination  of global  cluster properties,  the application  of the
Tully-Fisher  and Fundamental  Plane  relations,  the completeness velocities of the resulting GCVFS, and our  background
subtraction procedure.

\subsection{The Sloan Digital Sky Survey}
\label{sec:EDR}
The Sloan Digital Sky  Survey \citep[SDSS;][]{York00} is a photometric
and  spectroscopic  survey  being  carried  out at  the  Apache  Point
Observatory using a 2.5-m telescope equipped with modern CCD detectors
\citep{Gunn98}.   This  ambitious project  will  eventually cover  the
Northern  Galactic Hemisphere,  almost one  quarter of  the  sky.  The
photometric  survey  provides  nearly  simultaneous  imaging  in  five
passbands  \citep[$u$,  $g$,  $r$,  $i$,  $z$;  ][]{Fukugita96}.   The
spectroscopic  survey  consists   of  medium  resolution  ($\lambda  /
\Delta\lambda$  =  2000)  spectra  of  galaxies,  quasars,  and  stars
selected systematically from the photometric catalogs.

The first  public release of SDSS  data occurred in June  2001, and is
known as the Early  Data Release \citep[EDR;][]{Stoughton02}.  The EDR
consists of imaging data  covering approximately 460 square degrees of
sky, as well as approximately 55,000 spectra.

The   photometric   and   spectroscopic   pipelines   \citep{Lupton01,
Stoughton02} measure numerous quantities that are available as part of
the  EDR.  Below  we describe  only those  which will  be used  in the
present analysis.  Note that because the photometric system has yet to
be  finalized, we  denote the  current, temporary  magnitudes  with
asterisks, while references to the bandpasses remain simply italicized.

The photometric pipeline computes two types of magnitudes appropriate for use on galaxies:  Petrosian and model.
Petrosian magnitudes attempt to measure a constant fraction of the
total light of a galaxy, independent of surface brightness, distance,
or photometric model.  The Petrosian radius is computed in the $r$
band, and Petrosian magnitudes in all bands are based upon the amount
of flux within two Petrosian radii.  High signal-to-noise measurements
of the Petrosian magnitude can be made for galaxies as faint as r$^*
\sim$ 20.  As Petrosian magnitudes become noisy, model magnitudes
provide a more robust estimate of the total light of a galaxy.  To
compute model magnitudes, the photometric pipeline fits a pure de
Vaucouleurs profile and a pure exponential profile to the two
dimensional $r$ image of each galaxy.  The best-fitting model is then
applied, up to an amplitude change, to images in the other bands.  All
fits account for the effects of local seeing.  Parameters derived from
these fits include {\tt r\_deV}, half the length of the major axis of
the ellipse that encloses half the light of the galaxy; and {\tt
ab\_exp} and {\tt ab\_deV}, the best-fitting minor-to-major axis
ratios for the two models.

A bug in the SDSS photometric pipeline causes the model magnitudes of
bright galaxies to be $\sim$0.2 magnitudes too bright.  The
inverse Fundamental Plane relation we use to estimate the circular velocities
of early type galaxies (\S{\ref{sec:early}}) was calibrated using model
magnitudes afflicted by the same bug.  We therefore continue in the
use of EDR model magnitudes for early types in order to remain
consistent with this calibration.  The inverse Tully-Fisher relation
we apply to late type galaxies (\ref{sec:late}) was calibrated using
magnitudes extrapolated to infinity assuming exponential surface
brightness profiles.  Correctly-calculated model magnitudes would be
the optimal choice for consistency with this calibration.  In their
absence, we use Petrosian magnitudes for late type galaxies.

We  adopt galactic extinction  corrections as functions  of position
based upon \citet{Schlegel98}, and available from the SDSS database as
the  {\tt reddening}  parameter.  All  magnitudes in  the remainder  of this
paper refer to reddening-corrected values.

The   spectroscopic   pipeline  determines   both   an  emission   and
cross-correlation redshift.  Final redshifts ({\tt z}) are assigned to
each  galaxy  by  choosing  the  method with  the  highest  confidence
interval.

\subsection{Cluster Sample}
\begin{table*}
\begin{minipage}{170mm}
\caption{Galaxy Cluster Sample}
\begin{tabular}{ccccccccccccc}
RA          & DEC         & $z$   & $\sigma_{{\rm cl}}$ & $\Delta\sigma_{{\rm cl}}$ & $R_{{\rm vir}}$ & $V_{{\rm compl}}$  & $\log_{10}N_{{\rm 200}}$ & $\Delta \log_{10}N_{{\rm 200}}$ & $\beta$ & $\Delta \beta$ & $N_{{\rm 200}}^\prime$ & $\Delta {N_{200}^\prime}$ \\
h:m:s     & \degr:\arcmin:\arcsec & & $\kms$  & $\kms$                & Mpc       &      $\kms$         &                 &                             &         &                     &                        &                                \\\hline
00:21:40.841  & -0:55:34.17 & 0.105 & 370 & 71 & 0.77 & 65 & -0.50 & 0.08 & -1.47 & 0.35 & -0.83 s& 0.04 \\
00:23:31.152  & -0:48:10.08 & 0.063 & 540 & 97 & 1.20 & 50 & -0.96 & 0.13 & -3.20 & 0.50 & -0.88 & 0.05 \\
00:29:14.952  & -0:08:42.36 & 0.060 & 716 & 109 & 1.59 & 47 & -0.58 & 0.06 & -3.08 & 0.28 & -0.53 & 0.04 \\
00:46:15.528  &  0:09:08.64 & 0.114 & 582 & 74 & 1.20 & 73 & -0.81 & 0.11 & -2.24 & 0.47 & -0.94 & 0.06 \\
00:47:20.400  & -0:53:37.15 & 0.117 & 660 & 117 & 1.36 & 76 & -0.52 & 0.06 & -3.52 & 0.40 & -0.46 & 0.04 \\
00:56:00.120  &  0:37:56.74 & 0.067 & 557 & 81 & 1.23 & 50 & -0.92 & 0.11 & -3.76 & 0.43 & -0.69 & 0.03 \\
00:57:25.752  & -0:31:12.00 & 0.044 & 465 & 40 & 1.06 & 38 & -0.68 & 0.06 & -1.96 & 0.18 & -1.01 & 0.02 \\
01:15:08.784  &  0:15:56.88 & 0.045 & 599 & 44 & 1.36 & 39 & -0.56 & 0.05 & -2.37 & 0.13 & -0.70 & 0.03 \\
01:16:39.816  &  0:37:25.68 & 0.044 & 601 & 48 & 1.37 & 37 & -1.19 & 0.14 & -3.42 & 0.36 & -0.93 & 0.02 \\
01:31:08.928  &  0:31:23.52 & 0.079 & 533 & 64 & 1.15 & 56 & -0.70 & 0.08 & -2.71 & 0.34 & -0.73 & 0.04 \\
01:34:49.824  & -0:36:39.24 & 0.081 & 553 & 81 & 1.19 & 57 & -0.75 & 0.10 & -1.34 & 0.53 & -0.98 & 0.11 \\
01:37:25.704  & -0:28:08.40 & 0.056 & 432 & 123 & 0.97 & 44 & -1.13 & 0.21 & -2.50 & 0.84 & -1.20 & 0.08 \\
02:02:15.600  & -0:56:36.96 & 0.042 & 310 & 44 & 0.71 & 37 & -1.10 & 0.09 & -3.21 & 0.22 & -0.93 & 0.01 \\
03:06:17.376  & -0:08:34.80 & 0.109 & 585 & 56 & 1.22 & 70 & -0.64 & 0.07 & -1.96 & 0.33 & -0.84 & 0.06 \\
03:26:23.232  & -0:38:51.72 & 0.037 & 324 & 56 & 0.74 & 35 & -1.08 & 0.13 & -2.36 & 0.32 & -1.26 & 0.01 \\
10:07:55.441  &  0:35:29.04 & 0.097 & 493 & 49 & 1.04 & 64 & -0.51 & 0.05 & -1.86 & 0.23 & -0.71 & 0.04 \\
10:23:32.161  &  0:10:25.32 & 0.095 & 526 & 73 & 1.11 & 65 & -0.73 & 0.19 & -1.17 & 1.05 & -1.17 & 0.12 \\
10:49:15.839  &  0:57:15.84 & 0.106 & 474 & 78 & 0.99 & 68 & -0.74 & 0.10 & -1.67 & 0.43 & -1.05 & 0.04 \\
10:50:04.559  &  0:21:06.12 & 0.039 & 281 & 31 & 0.64 & 35 & -1.07 & 0.13 & -2.15 & 0.34 & -1.35 & 0.02 \\
12:46:54.240  &  0:18:31.32 & 0.089 & 844 & 89 & 1.80 & 62 & -0.50 & 0.06 & -2.46 & 0.28 & -0.57 & 0.06 \\
12:47:43.440  & -0:09:07.20 & 0.088 & 994 & 102 & 2.13 & 61 & -0.73 & 0.11 & -2.94 & 0.48 & -0.74 & 0.10 \\
13:19:14.641  & -0:53:58.20 & 0.083 & 704 & 73 & 1.52 & 58 & -0.45 & 0.04 & -3.44 & 0.16 & -0.31 & 0.02 \\
13:21:18.241  & -0:44:19.32 & 0.108 & 873 & 131 & 1.82 & 70 & -0.52 & 0.07 & -2.85 & 0.33 & -0.54 & 0.06 \\
13:24:22.321  &  0:46:55.92 & 0.108 & 575 & 129 & 1.20 & 70 & -1.06 & 0.14 & -4.30 & 0.53 & -0.76 & 0.04 \\
13:45:05.760  &  0:13:55.20 & 0.089 & 284 & 43 & 0.61 & 62 & -0.89 & 0.14 & -2.86 & 0.55 & -0.88 & 0.04 \\
13:45:15.601  & -0:08:30.84 & 0.088 & 362 & 42 & 0.77 & 64 & -1.28 & 0.31 & -3.50 & 1.19 & -1.13 & 0.08 \\
14:14:41.279  & -0:24:20.44 & 0.137 & 761 & 80 & 1.52 & 81 & -0.04 & 0.02 & -2.00 & 0.17 & -0.11 & 0.03 \\
14:15:05.039  & -0:29:08.16 & 0.141 & 930 & 113 & 1.85 & 84 & -0.08 & 0.03 & -2.07 & 0.23 & -0.13 & 0.04 \\
14:29:36.962  &  0:24:36.00 & 0.055 & 281 & 36 & 0.63 & 44 & -1.05 & 0.15 & -1.74 & 0.40 & -1.48 & 0.04 \\
15:08:31.919  & -0:16:03.00 & 0.090 & 737 & 61 & 1.57 & 65 & -0.42 & 0.05 & -2.07 & 0.26 & -0.54 & 0.05 \\
15:11:19.918  & -0:02:26.16 & 0.091 & 455 & 49 & 0.97 & 62 & -0.58 & 0.06 & -1.63 & 0.27 & -0.76 & 0.07 \\
15:16:23.522  &  0:06:09.00 & 0.119 & 675 & 120 & 1.39 & 76 & -0.38 & 0.04 & -1.24 & 0.34 & -0.47 & 0.07 \\
15:17:25.920  & -0:38:24.36 & 0.116 & 758 & 92 & 1.56 & 77 & -0.22 & 0.03 & -2.94 & 0.21 & -0.22 & 0.03 \\
15:29:12.719  & -0:15:07.92 & 0.087 & 459 & 54 & 0.98 & 63 & -0.62 & 0.07 & -2.52 & 0.27 & -0.72 & 0.04 \\
\label{datatable}
\end{tabular}
RA and DEC are both in J2000.0; $z$ is the cluster redshift;
$\sigma_{{\rm cl}}$ is the line-of-sight cluster velocity dispersion,
as described in \S{\ref{sec:global}}; $\Delta\sigma_{{\rm cl}}$ is the
bootstrap error on $\sigma_{{\rm cl}}$; $R_{vir}$ is the virial radius
of the cluster, as determined from Equation \ref{rvir}; $V_{compl}$ is
the completeness limit of the cluster GCVF, computed as described in
\S{\ref{sec:completeness}}; $N_{{\rm 200}}$ and $\beta$ and their
errors, $\Delta N_{{\rm 200}}$ and $\Delta \beta$ are found by fitting
Equation \ref{clusterfit} to the cluster data; $N_{{\rm 200}}^\prime$
and the associated error, $\Delta N_{{\rm 200}}^\prime$, is found by
fitting Equation \ref{clusterfit} while fixing $\beta$ to its average
value of -2.5.  
\end{minipage}
\end{table*}

Several cluster catalogs have been constructed from SDSS data, using a
variety of identification algorithms \citep{Bahcall03}.  In order to
compute accurate GCVFs down to the lowest possible circular
velocities, we used the nearest clusters.  Miller \etal,
2002 (in prep) have identified 62 clusters in the Equatorial Stripes
of the SDSS using the C4 algorithm \citep{Nichol01, Gomez02}.  This
algorithm uses the observed phenomenon that galaxies in clusters have
similar colors as well as locations.  For each galaxy, the algorithm
determines the number of neighbors in the seven-dimensional space
spanning RA, DEC, redshift, and the four SDSS colors.  This number of
neighbors is compared to the distribution found for 100 random field
galaxies.  Based upon this comparison, the probability that the galaxy
is a field galaxy is computed.  Galaxies that are probably not field
galaxies are identified using the False Discovery Rate thresholding
technique \citep{Miller01}.

In order to create as large a sample as possible, we supplement the C4
clusters with 51 nearby NED\footnote{The NASA/IPAC Extragalactic
Database (NED) is operated by the Jet Propulsion Laboratory,
California Institute of Technology, under contract with the National
Aeronautics and Space Administration.} clusters in the same area.

After removing 18 duplicates between the C4 and NED catalogs, 5
clusters with virial radii which extend beyond the SDSS data limits
(\S{\ref{sec:global}}), 36 clusters (mostly NED) with fewer than 20
spectroscopically confirmed members (\S{\ref{sec:global}}), and 20
clusters with data defects or that were not rich enough to yield high
signal-to-noise GCVFs, we are left with 34 clusters, which are listed
in Table \ref{datatable}.

\subsection{Determination of Global Cluster Properties}
\label{sec:global}

As discussed in \S{\ref{sec:intro}}, the GCVF may depend on
environment, and thus on cluster mass.  While a reliable cluster mass
estimate is difficult to obtain, the line-of-sight velocity
dispersion, $\sigma_{{\rm cl}}$, is relatively straightforward to
measure, and can be used as a proxy for the mass.  We calculate
$\sigma_{{\rm cl}}$ using the following procedure.

\begin{figure}
\psfig{figure=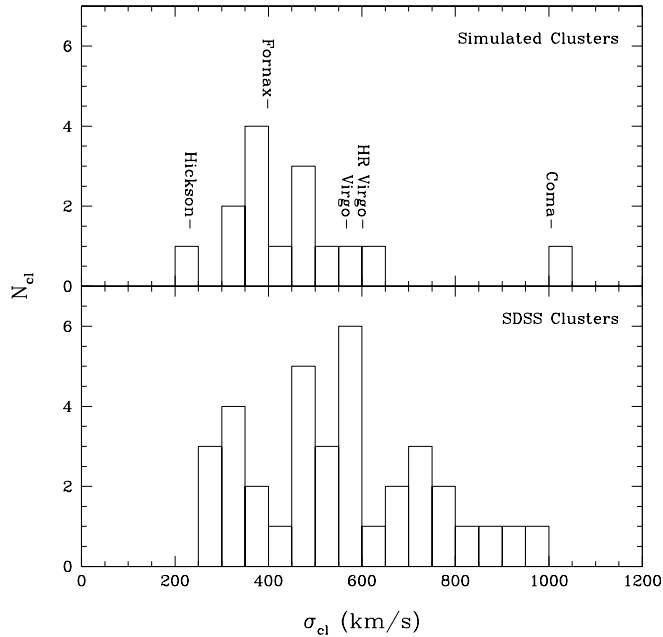,width=3.5truein,height=3.5truein}
\caption{{\em Top Panel:}  The distribution of cluster velocity
  dispersions for the fifteen simulations
  described in \S{\ref{sec:sims}}.  {\em Bottom Panel}  The
  distribution of cluster velocity dispersions, calculated by the
  methods described in \S{\ref{sec:global}}, for our sample of 34
  SDSS clusters.}
\label{sigmaplot} 
\end{figure}

All galaxies with spectroscopic redshifts within a radius of either 1.5 Mpc
or the largest radius allowed by the extent of the data, whichever was
smaller, were selected.  Galaxies were ordered by recessional
velocity, and gaps larger than 800 $\kms$ were used to remove
interlopers.  A robust weighted gap estimator \citep{Beers90} was used
to determine $\sigma_{{\rm cl}}$ from the remaining velocities.  This
process was performed iteratively until the velocity dispersion
stabilized.  The resulting velocity dispersion was divided by a factor
of $(1+z)$ to correct it to the cluster frame \citep{Danese80}.  In
order to ensure accurate estimates of $\sigma_{{\rm cl}}$, we rejected
36 clusters with fewer than 20 confirmed members.  Errors were
estimated using bootstrap re-sampling, and are typically 10--20 per
cent.  The velocity dispersions and errors for each cluster can be found in Table
\ref{datatable}, columns 4 and 5, and the bottom panel of Figure
\ref{sigmaplot} shows the distribution of $\sigma_{{\rm cl}}$.

In order to compare our circular velocity functions to each other and
to models, we calculated them within a consistent aperture.  The
virial radius, $R_{{\rm vir}}$, is a natural aperture to select.
\citet{Girardi01} present the following empirical relation between
$R_{{\rm vir}}$ and $\sigma_{{\rm cl}}$:

\begin{equation}
R_{{\rm vir}}    \sim   \frac{0.0017    \sigma}{(1    +   z)^{3/2}}
\mathrm{~~(km^{-1} ~s}~h_{100}^{-1} ~\mathrm{Mpc)}.
\label{rvir}
\end{equation}

\noindent Here R$_{{\rm vir}}$ is the radius within which the average
density is equal to $\Delta \sim 200$ times the critical density.
Virial radii computed from Equation \ref{rvir} are listed in column 6
of Table \ref{datatable}.  We did not recalculate $\sigma_{{\rm cl}}$
using galaxies only within the virial radius for two reasons.  First,
for many of the clusters, nearly all spectroscopically confirmed
cluster members were contained within $R_{{\rm vir}}$. Second, for the
remainder of clusters, we wished to determine $\sigma_{{\rm cl}}$ using
as many redshifts as possible.

\subsection{Determination of Galaxy Circular Velocities}
\label{sec:vcirc}

Each  galaxy  within  R$_{{\rm vir}}$   and  with  $r^*$  $<$  21.5
\citep[the  star/galaxy separation limit;  ][]{Lupton01} is  given the
designation  of  Early  (E,  S0,  Sa)  or Late  (Sb,  Sc,  Irr)  type.
\citet{Strateva01}  find   that  $u^*-r^*  <  2.22$   is  optimal  for
separating these classes.  For selecting early types, the completeness
is 68 per  cent, while the reliability is 81  per cent.  For selecting
late types, the completeness is 77 per cent, and the reliability is 96
per cent.   We then  translated the surface  photometry of  each galaxy
into a circular velocity  using the conversions described in the
following two subsections.

\subsubsection{Late Type Galaxies}
\label{sec:late}

Disk galaxies exhibit a tight correlation between luminosity and
circular velocity.  The Tully-Fisher relation (TFR) allows one to
compute the former given the latter.  We require the inverse TFR,
which converts from luminosity to circular velocity.  Neither the TFR
nor the inverse TFR have been calibrated in SDSS passbands.  We
therefore follow \citet{Sheth03} in adopting the inverse I-band
relation presented by \citet{Giovanelli97} for 360 spiral galaxies
deemed to be members of 24 clusters (their ``{\bf in}'' sample).
Although it was calibrated using cluster data, we applied the
following relationship to late type galaxies in the determination of
both the field and cluster GCVFs:

\begin{equation}
\log_{10}(2V_{{\rm c}}) - 2.5 = -\frac{21.10}{7.94} - \frac{(M_I - 5\log_{10} h_{100})}{7.94},
\label{TFR}
\end{equation}

\noindent where M$_I$ is the {\em k}-corrected, extinction-corrected
I-band absolute magnitude, extrapolated to infinity assuming an
exponential surface brightness profile; and $V_{{\rm c}}$ is an
estimate of the maximum rotational velocity of a galaxy.  Although
disk galaxy rotation curves exhibit a variety of shapes, the circular
velocity generally increases with radius until it reaches a broad peak
at $\sim$10 kpc, beyond which it remains relatively constant.  Thus
$V_{{\rm c}}$ should be similar to the circular velocity
measured at the flat part of the rotation curve.

In order to convert the observed SDSS magnitudes into those upon which
the TFR  is based,  we first computed  the {\it r}-band  absolute total
magnitude of each late type galaxy:

\begin{equation}
M_{r^*} = m_{r^*} - 5\log d_L(z) + 5 - K(z),
\end{equation}

\noindent where $m_{r^*}$ is the apparent {\it r}-band Petrosian
magnitude, $d_L$ is the luminosity distance in parsecs, and $K(z)$ is
the {\em k}-correction, computed using {\sc kcorrect v1\_11}
\citep{Blanton03}.  We then use the conversion from \citet{Fukugita95}
for late type galaxies: $M_{r^*}-M_I\sim0.9$.  

Internal extinction
corrections were estimated using the procedure of \citet{Tully85}:

\begin{equation}
A           =          -2.5\log\left[f\left(1+e^{-a}\right)          +
\left(1-2f\right)\left(\frac{1-e^{-a}}{a}\right)\right],
\end{equation}

\noindent where $f = 0.1$ is the fraction of stars homogeneously mixed
with a dust layer having opacity $\tau = 0.28$; $a = \tau \sec i$; and
$i$ is the inclination, estimated as:

\begin{equation}
\cos^2(i) = {\frac{q^2 - q_0^2}{1 - q_0^2}}.
\end{equation}

\noindent Here $q_0  = 0.20$ is the intrinsic thickness  and $q = {\tt
ab\_exp}$  is  the  {\it   r}-band  axis  ratio.   The  extinction  is
considered to be constant for $i>80\degr$.

The internal extinction corrections were on the order of $\sim$0.5 magnitudes.

\subsubsection{Early Type Galaxies}
\label{sec:early}

Early type galaxies occupy a thin plane in the space defined by
stellar velocity dispersion, physical size, and surface brightness.
Given the velocity dispersion and surface brightness, one can use the
Fundamental Plane relation (FP; \citet{Djorgovski87, Dressler87}) to estimate the size.  In
order to construct the GCVF from SDSS photometry, we require a
relation that will yield the velocity dispersion (which is simply
related to the circular velocity) given a size and surface brightness.

In a series of three papers \citep{Berna03, Bernb03, Bernc03}, Bernardi and collaborators constructed a sample of $\sim$9000 SDSS early type
galaxies, from which they computed the joint distributions of
$\sigma_{{\rm los}}$, the line-of-sight velocity dispersion measured
through a circular aperture with a radius equal to one-eighth the
half-light radius ($r_{{\rm 0}} =
\mathtt{r\_dev}\sqrt{\mathtt{ab\_dev}}$); $I_0$, the average surface
brightness within $r_0$; and $R_{{\rm 0}}$, the half-light radius in
kpc.  In order to determine the coefficients of the FP relation,
\citet{Bernc03} used the above joint distributions to minimize the
sum of $\Delta_1 = \log_{10} R_0 - a \log_{10}\sigma_{{\rm los}} - b
\log_{10} I_0 - c$ over all galaxies in their sample.  In order to compute the
coefficients for the inverse FP relation,

\begin{equation}
\log_{10}   \sigma_{{\rm los}}  =   a^\prime  \log_{10}   R_{{\rm 0}}  +   b^\prime  \log_{10}   I_0  + c^\prime, 
\label{FP}
\end{equation}

\noindent we used the same distributions to minimize the sum of
$\Delta_{{\rm invFP}}^2=\left(\log_{10}\sigma_{{\rm los}} - a^\prime\log_{10}
R_{{\rm 0}} - b^\prime\log_{10} I_0 - c^\prime\right)^2$ over all galaxies.
We found $a^\prime=0.597$, $b^\prime=-0.447$, and $c^\prime=5.449$.

For each early type galaxy, we estimated $\sigma_{{\rm los}}$ using
Equation \ref{FP}.  Following \citet{Berna03}, we set

\begin{equation}
\mu_0 = M_{r^*} + 2.5 \log(2\pi r_0^2) - K(z) - 10 \log(1+z),
\end{equation}

\noindent  where $\mu_0$  =  -2.5 $\log$  I$_0$  and K(z)  is as
previously defined.

In determining R$_0$, the physical characteristic size, we first
measured the angular characteristic size for each galaxy, $r_{{\rm
0}}$.  Because the inverse FP that we used was calibrated against
galaxies at a variety of redshifts, the angular sizes measured in each
band were linearly interpolated in order to determine the angular size
at a rest wavelength corresponding to the central wavelength of each
filter.

In order to plot early type galaxies on the same velocity scale as
late types, we converted $\sigma_{{\rm los}}$ to $V_{{\rm c}}$ (see
\S{\ref{sec:late}}).  Fitting non-parametric, spherical models to line
profile and radial velocity dispersion data for 21 elliptical
galaxies, \citet{Kronawitter00} found that the circular velocity
profiles of elliptical galaxies are flat to the 10 per cent level
outside a radius of 0.3 effective radii.  \citet{Gerhard01} showed that
there is a tight relation between the circular velocities in the flat
regions of the rotation curves and the central velocity dispersions
for these ellipticals.  \citet{Ferrarese02} examined this
relation, taking central velocity dispersions from \citet{Davies87}
for the ellipticals and correcting them to an aperture of one-eighth
the effective radius to find a relation between $\log_{10}V_{{\rm c}}$
and $ \log_{10}\sigma_{{\rm los}}$.  For an isothermal sphere, $V_{{\rm
c}} = \sqrt{2} \sigma_{{\rm los}}$.  Using the same data as
\citet{Ferrarese02}, we found that the isothermal assumption
systematically underpredicts $V_{{\rm c}}$, but that the data are
consistent with the following relation:

\begin{equation}
V_{{\rm c}} \approx 1.54 \sigma_{{\rm los}}.
\label{isothermal}
\end{equation}

\noindent For each observed early type galaxy, we use Equation \ref{isothermal}
to determine $V_{{\rm c}}$.

\subsubsection{Accounting for Scatter in the Derived Circular Velocities}
\label{sec:scatter}

Values of $V_{{\rm c}}$ for individual galaxies were computed with the
mean relations in \S{\ref{sec:early}} and
\S{\ref{sec:late}}. \citet{Sheth03} note that if the scatter is
significant, the use of mean relations can significantly distort the
shape of the GCVF.  Therefore, in constructing our GCVFs, we accounted
for the scatter by representing each galaxy's $V_{{\rm c}}$ as a
probability distribution, rather than a single value.  Because we
calculated $V_{{\rm c}}$ using different relations for early and late
type galaxies, this distribution depends on galaxy type.

The circular velocities for late type galaxies were computed directly
from the inverse TFR (\S{\ref{sec:late}}).  We therefore adopted the model
for the scatter around $V_{{\rm c}}$ provided by \citet{Giovanelli97}.

\begin{equation}
<\Delta_{{\rm inv TF}}^2> = \frac{(-0.28(\log_{10}(2V_{{\rm c}}) - 2.5) + 0.26)^2}{63.04}.
\label{TFscatter}
\end{equation}

The circular velocities for early type galaxies were computed using
Equations \ref{FP} and \ref{isothermal}.  There were two sources of
scatter in determining $V_{{\rm c}}$: the scatter around $\sigma_{{\rm
los}}$ and the uncertainty in the relationship between $\sigma_{{\rm
los}}$ and $V_{{\rm c}}$.  We estimated the former by exchanging V and
R in the expression for $<\Delta_1^2>$ in Equation 4 of
\citet{Bernc03} and using the joint distributions of $\sigma_{{\rm
los}}$, $R_{{\rm 0}}$, and $I_{{\rm 0}}$ presented in their Table 1.
Doing so yielded a scatter of $<\Delta_{{\rm invFP}}^2> = 0.061$ in
$\log_{10}\sigma_{{\rm los}}$.  \citet{Sheth03} found that the scatter
in $\log_{10}\sigma_{{\rm los}}$ depends slightly on
$\log_{10}\sigma_{{\rm los}}$.  This was a small effect
compared to our other sources of error; we therefore treated the scatter in the
Fundamental Plane as constant.  The conversion from $\sigma_{{\rm
los}}$ to $V_{{\rm c}}$ (Equation \ref{isothermal}) was
calibrated on only $\sim$20 elliptical galaxies.  Therefore a robust
model of the scatter due to this conversion was not possible and we neglected it.

Because the Fundamental Plane and Tully-Fisher relations were
calibrated using samples created under more stringent selection
criteria than we used, the above relations underestimate the
scatter that exists in our data.
  
\subsection{Background Subtraction}

The largest  uncertainty in our determinations of  cluster GCVFs stemmed
from  the difficulty  in  determining cluster  membership.  Because  a
spectrum  is not  available  for every  galaxy  in the  field of  each
cluster,   our  background  subtraction   is  statistical.    We  calculated the  circular velocity for  every galaxy within  the virial
radius  of each  cluster  under the  assumption  that it  lies at  the
cluster redshift.  Clearly, some  galaxies have redshifts greater
or less than that of  the cluster, and our calculations were incorrect
for  these galaxies.   This effect  was removed  during  our background
subtraction procedure. Again using  the cluster redshift, we calculated
the  circular velocity  for every  galaxy within  an  annulus
bounded by inner and outer radii  of 3 and 5 Mpc, respectively.  These
galaxies should  have a redshift  distribution similar to that  of the
cluster   interlopers.   Thus,   by   subtracting  the   appropriately
normalized GCVF of galaxies in  the annulus, we corrected for the
interlopers.

The error in the number of galaxies in each circular velocity bin are
calculated assuming Poisson noise in both the total counts within the
virial radius, $C_{{\rm tot}}$, and the counts within the background
annulus, $C_{{\rm background}}$:

\begin{equation}
\sigma_N^2 = C_{{\rm tot}} + f^2 C_{{\rm background}},
\end{equation}

\noindent where $f$ is the normalization factor applied to the
background annulus counts to account for the difference in its area
and that of the virialized part of the cluster ($0.02 \la f \la 0.3$).  

The contribution to the total error in the GCVFs from the Poisson error in the background counts decreases with increasing $V_{{\rm c}}$.
Above $\sim$200 $\kms$, the contribution is less than 40 per cent.
  
\subsection{Completeness}
\label{sec:completeness}

\begin{figure}
\psfig{figure=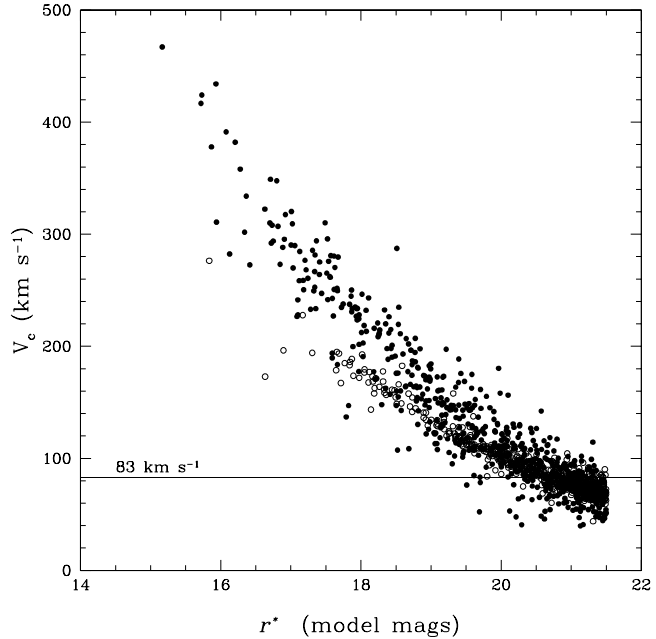,width=3.5truein,height=3.5truein}
\caption{Illustration of the completeness limit for one cluster in our
  sample.  Each point represents a galaxy within $R_{vir}$ for a
  representative cluster in our SDSS sample.  No background
  subtraction has been carried out.  Filled circles are early type galaxies, and hollow circles denote late type galaxies.  Ninety per cent of all galaxies
  with $21 > {\rm r}^* > 21.5$ lie below the completeness velocity of
  $V_{{\rm compl}} = 83 \kms$ for this cluster.}
\label{completeness} 
\end{figure}

Because our procedure distinguishes between morphological types, we could
assign different values of $V_{{\rm c}}$ to galaxies of the same
luminosity.  Our requirement that $r^*<21.5$ therefore leads to a cut
in $V_{{\rm c}}$ that depends on both redshift and galaxy type.  Figure
\ref{completeness} illustrates the effect our magnitude limit has on
$V_{{\rm c}}$ for a representative cluster.  The filled points represent
early type galaxies, while the hollow points represent late type
galaxies.  For this cluster, 90 per cent of galaxies with $21 < {\rm
r}^* < 21.5$ have $V_{{\rm c}} > 83 \kms$.  Below this completeness
velocity, $V_{{\rm compl}}$, significant numbers of galaxies are
missing from our sample.  All GCVFs in what follows are truncated
below $V_{{\rm compl}}$, which was calculated independently for each
cluster.  The adopted values are listed in Table \ref{datatable},
column 7.

\section{Methods:  Simulations}
\label{sec:sims}

\begin{table}
\caption{Simulation Sample}
\begin{tabular}{ccccccc}
              & $\sigma_{{\rm cl}}$          & M$_{200}$              & R$_{200}$       & m$_{DM}$              & N$_{DM}$ & $\epsilon$   \\
              & $\kms$                       & $10^{13} M_{\sun}$     & $h_{70}$Mpc             & $10^{8} M_{\sun}$     & $10^6$   & kpc         \\\hline
Box01         & 525                          & 29.4                   & 1.73            & 1.9                   & 1.5      & 7.14         \\
Box02         & 458                          & 26.6                   & 1.67            & 1.9                   & 1.4      & 7.14         \\
Box03         & 383                          & 21.7                   & 1.56            & 1.9                   & 1.1      & 7.14         \\
Box04         & 456                          & 18.4                   & 1.48            & 1.9                   & 0.97       & 7.14         \\
Box05         & 451                          & 17.9                   & 1.46            & 1.9                   & 0.94       & 7.14         \\
Box06         & 342                          & 15.4                   & 1.39            & 1.9                   & 0.81       & 7.14         \\
Box07         & 388                          & 13.0                   & 1.32            & 1.9                   & 0.68       & 7.14         \\
Box08         & 332                          & 11.6                   & 1.27            & 1.9                   & 0.61       & 7.14         \\
Box08         & 369                          & 10.3                   & 1.22            & 1.9                   & 0.54       & 7.14         \\
Box10         & 366                          & 8.9                    & 1.26            & 1.9                   & 0.47       & 7.14         \\\hline
Hickson       & 234                          & 2.5                    & 0.76            & 4.41                  & 0.0486     & 2.5          \\
Fornax        & 400                          & 5.9                    & 0.65            & 4.41                  & 0.118     & 2.5          \\
Virgo         & 565                          & 30.4                   & 1.75            & 14.9                  & 0.179     & 1.25          \\
Coma          & 1007                         & 133.6                  & 2.86            & 119.2                 & 0.090      & 3.75          \\\hline
HR Virgo  & 600                          & 30                     & 1.75            & 3                     & 1.00      & 1.25         \\

\label{simtable}
\end{tabular}
Further details on the simulated clusters named Box01 -- Box10 can be found in
\citet{Reed03}.  The simulations labelled Hickson, Fornax, Virgo, and Coma are described in \citet{Borgani02}.
The last is used in \S{\ref{sec:hires}}. {\em Columns:} $\sigma_{{\rm
cl}}$ is the velocity dispersion of the simulated cluster, $R_{200}$
is the radius within which the average density is 200 times the
critical density, $M_{200}$ is the mass within $R_{200}$, $m_{DM}$ is
the mass of each dark matter particle, $N_{DM}$ is the number of dark
matter particles within $R_{200}$, and $\epsilon$ is the force
softening.
\end{table}

In order to compare our observational results against simulations, we
compiled a set of 15 simulations of the dark matter components of
clusters.  Each simulated cluster was evolved in the currently favored
$\Lambda$CDM cosmology with $\Lambda$=0.7, H$_0$ = 70 $\kmsmpc$,
$\Omega_0=0.3$, and $\sigma_8 = 1$.  Ten are the most massive clusters
from the zero redshift output of the 50$^3$ $h_{100}^{-3}$ Mpc$^3$
simulation described in \citet{Reed03}.  This simulation has a
particle mass of 1.9$\times$10$^{8}$M$_{\sun}$ and a force resolution
of 7.14 kpc.  An additional four clusters were simulated using the
volume renormalization technique \citep{Katz93} and have
(0.5-0.9)$\times$10$^5$ particles within the virial radius and force
softenings of $\la$1 per cent of the virial radius.  Further details
on these four simulations can be found in \citet{Borgani02}.

The 14 cluster simulations were of sufficient resolution to
robustly determine peak circular velocities in subhalos above the
completeness velocities of our observed GCVFs (see
\S{\ref{sec:completeness}}).  They were of insufficient resolution,
however, to accurately trace subhalo internal density profiles.  These
were necessary to determine the effects of baryons on galaxy rotation
curves, as we describe in \S{\ref{sec:hires}}.  For this purpose, we used a
very high resolution {\it N}--body simulation of a galaxy cluster with
$\sigma_{{\rm cl}} = 600\kms$ and $M_{{\rm vir}} = 3\times
10{^{14}}M_{\sun}$,  represented by approximately one million particles
within $R_{{\rm vir}}$ = 1.75 Mpc.  It was evolved in a
$100^3$ Mpc$^3$ volume where the force spline softening was 1.25 kpc
and the large scale structure was sampled at a lower resolution.

Subhalos were identified using {\sc SKID} \citep{Stadel01,
Governato97}, a halo finder based on local density maxima and
particulary suited to finding subhalos within larger structures.  Peak
circular velocities were also output by {\sc SKID}.  We used {\tt
Tipsy}\footnote{{\tt
http://www-hpcc.astro.washington.edu/tools/tipsy/tipsy.html}} to
extract full density profiles from our highest resolution simulation.

\section{Results}
\label{sec:results}

\subsection{The Shape of the GCVF}
\label{sec:shape}

\begin{figure}
\psfig{figure=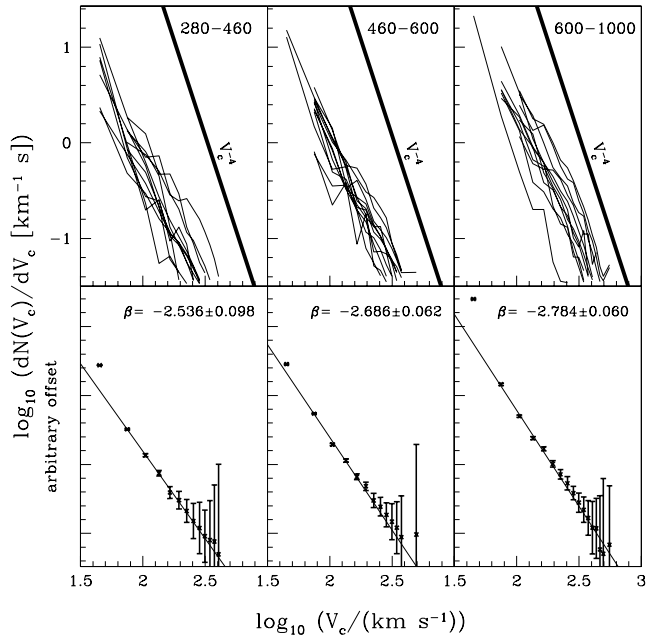,width=3.5truein,height=3.5truein}
\caption{The observed CGVF, binned by cluster velocity dispersion.
  {\em Top panels:} Individual cluster GCVFs are shown with thin
  lines.  The dark line in each panel has a slope of -4, and is
  provided for reference.  Cluster velocity dispersion ranges are
  shown in the upper right of each panel, in $\kms$.  {\em Bottom
  panels:} Each panel shows the composite functions constructed from
  clusters having $\sigma_{{\rm cl}}$ in the range indicated in the
  panel directly above.  Error bars take into account Poisson error in
  counting galaxies both within $R_{{\rm vir}}$ and in the background
  annulus.  A power law (Equation \ref{clusterfit}) was fitted to
  these data, and is shown as a thin solid line.  The slopes of the
  composite functions are shown in the upper right hand corner of each
  panel.}
\label{sigmabins} 
\end{figure}

\begin{figure}
\psfig{figure=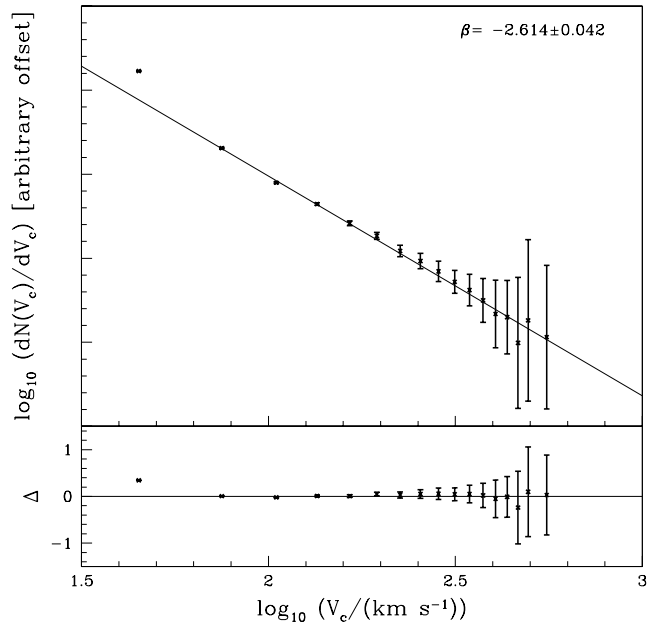,width=3.5truein,height=3.5truein}
\caption{The composite CGVF, constructed from all 34 SDSS clusters in
  our sample.  {\em Top panel:} The error bars were computed as in
  Figure \ref{sigmabins}.  The best-fitting power law is shown as a
  thin solid line, and its slope is shown in the upper right hand
  corner.  {\em Bottom panel:} The residuals of the composite cluster
  GCVF from the best fit power law.  The power law is a very good fit
  to the composite GCVF, except for the lowest $V_{{\rm c}}$ bin.}
\label{composite}
\end{figure}

A Press-Schechter mass function combined with $M \propto V^3$ and
a value of $n=2$ for the fluctuation-spectrum index give

\begin{equation}
\frac{dN\left(V_{{\rm pk}}\right)}{dV_{{\rm pk}}} \propto V_{{\rm pk}}^{-4}.
\end{equation}

\citet{Moore99} and \citet{Ghigna00} found the GCVF in high resolution
simulations of a Virgo-like cluster to be consistent with the above
power law.  \citet{Moore99} also found the observed GCVF of Virgo to
be consistent with it.  In Figure \ref{sigmabins} we plot the GCVFs
for our sample of 34 clusters in three cluster velocity dispersion
bins.  The bins were chosen such that each contains approximately the
same number of clusters.  The bold lines have a slope of -4 and are
for reference only.  From this figure, we can see that the observed
cluster GCVF is a power law.  The power law form of the cluster GCVF,
while predicted analytically and numerically, is nonetheless somewhat
surprising.  The cluster luminosity function is usually described by a
Schechter function.  Because the luminosities of both early and late type
galaxies are related to their circular velocities by power laws ($L
\propto V^n$), it is not obvious that the galaxy population as a whole
should obey both a Schecter luminosity function and a power law GCVF.

Key to the explanation is the fact that both early
and late types contribute to the GCVF, and that they have both
different relations between $M_{r^*}$ and $V_{{\rm c}}$, and make up
different fractions of the population with magnitude.  Because the
slope of the $V_{{\rm c}}-M_{{\rm r}}$ relation for early types is steeper
than for late types, late type galaxies were mapped onto a smaller
range of velocities than early types.  Thus, late type galaxies that
populate the knee in the luminosity function were mapped to lower
velocities than early type galaxies with the same luminosity.  The
proportion of early and late type galaxies that make up the knee in
the luminosity function affects the placement and prominence of the
knee in the GCVF.  In general, it was smeared out in the GCVF.
This is the main reason that our observed cluster GCVFs look like
power laws.  If a single luminosity-velocity relation is used to
compute a GCVF from a population of galaxies drawn from a Schechter
function, the resulting GCVF will also be a Schechter function
\citep{Cole89}.  If two different luminosity-velocity relations are
used in the computation, it is still possible to arrive at a Schechter
GCVF, as found by \citet{Gonzalez00} and \citet{Kochanek01}.  A
Schechter GCVF will not {\em necessarily} result, however, depending
upon the relative abundances of early and late type galaxies as a
function of luminosity.

Another reason for the power law GCVF is that our completeness limits
did not allow us to observe the shape of the GCVF at velocities
dominated solely by galaxies from the flat part of the luminosity
function, where we would expect the GCVF to flatten out.  Thus, very
few bins define the knee.

Finally, the signal-to-noise in each cluster GCVF is relatively low
due to the small number of galaxies used.  It would be difficult to
trace the high velocity exponential drop-off if the GCVF did follow a
Schechter function.  

In order to increase the signal to noise in in our determination of
the cluster GCVF, we built composite cluster GCVFs using a technique
analogous to that of \citet{Colless89} for building composite cluster
luminosity functions.  The number of galaxies in the $j$th $V_{{\rm
c}}$ bin of the composite cluster GCVF is given by:

\begin{equation}
N_{{\rm c}j} = \frac{N_{c0}}{m_j}\sum_{i}\frac{N_{ij}}{N_{i0}},
\end{equation}

\noindent where $N_{ij}$ is the number of galaxies in the $j$th bin of
the $i$th cluster GCVF, $N_{i0}$ is the normalization used for the
$ith$ cluster GCVF (taken as the number of galaxies brighter than
100$\kms$), $m_j$ is the number of clusters contributing to the $j$th
bin, and $N_{{\rm c0}}$ is the sum of all the normalizations:

\begin{equation}
N_{{\rm c0}} = \sum_{i}N_{i0}.
\end{equation}

\noindent The error on $N_{{\rm cj}}$ is:

\begin{equation}
\Delta N_{{\rm cj}} = \frac{N_{{c0}}}{m_j}\left[ \sum_i \left( \frac{\Delta N_{ij}}{N_{i0}}\right)^2 \right]^{\frac{1}{2}},
\end{equation}

\noindent where $\Delta N_{ij}$ is the error on the $j$th bin of the
$i$th cluster.
 
The composite cluster GCVFs in the three $\sigma_{{\rm cl}}$ bins are
shown in the bottom panel of Figure \ref{sigmabins}.  The composite
cluster GCVF constructed using all 34 clusters is shown in Figure
\ref{composite}.  An increase in signal to noise still yielded a power
law form.

\subsection{Is cluster substructure self-similar?}
\label{sec:selfsim}

\begin{figure}
\psfig{figure=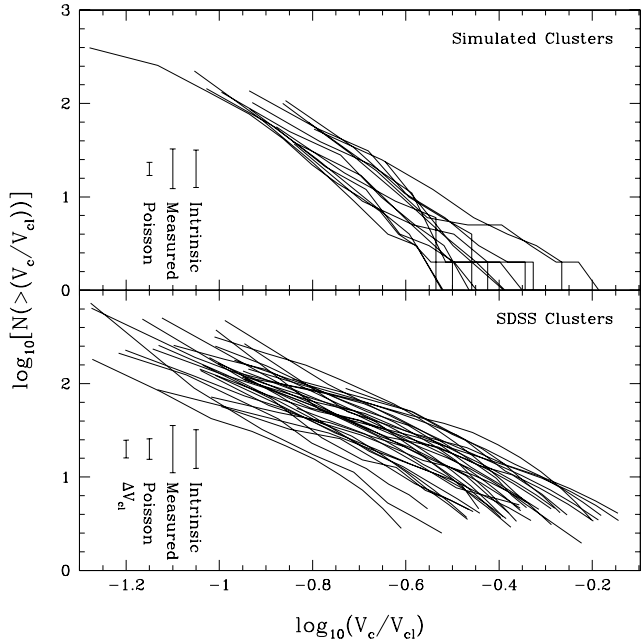,width=3.5truein,height=3.5truein}
\caption{The simulated and observed SDSS CGVFs, with galaxy circular
velocities scaled by cluster circular velocites.  The error bars have
length 2$\Delta$, where $\Delta$ is the RMS scatter at
$\log_{10}(V_{{\rm c}}/V_{{\rm cl}} \approx -0.7)$ due to various
sources, as labelled.  {\em Poisson} refers to Poisson noise in
counting galaxies within $R_{{\rm vir}}$ for the simulated clusters and
to Poisson noise in counting galaxies within $R_{{\rm vir}}$ and the
background annulus for the SDSS clusters.  {\em Measured} refers to
the total measured scatter.  {\em $\Delta V_{{\rm cl}}$} refers to the
scatter in $\log_{10}[N(>(V_{{\rm c}}/V_{{\rm cl}}))]$ due to errors
in measuring $V_{{\rm cl}}$, and {\em Intrinsic} refers to the portion
of the total measured scatter unaccounted for by Poisson scatter in
the case of the simulations or by both Poisson scatter and that
induced by $\Delta V_{{\rm cl}}$ for the SDSS clusters.}
\label{scaledclusters} 
\end{figure}

\citet{Moore99} found that the cumulative GCVF for a Virgo-like
simulation is identical to that of subhalos
within a Milky-Way-like simulation, provided the velocities are
scaled by the velocity of the parent halo ($V_{{\rm cl}} = {\sqrt
2}\sigma_{{\rm cl}}$).  In the top panel of Figure
\ref{scaledclusters}, we have plotted the cumulative, scaled GCVF for
all fifteen simulated clusters from Table \ref{simtable}.  Our
simulated clusters have very similar cumulative, scaled GCVFs, in agreement with the results
of \citet{Moore99}.  

In the bottom panel of Figure \ref{scaledclusters}, we plot the same
quantity for our sample of SDSS clusters.  The slopes and offsets of
the simulated and observed clusters cannot be compared directly, since
the effects of baryons have not been included in the simulations.
However, assuming that baryonic physics changes the shapes of all the
GCVFs in a similar way, we can compare the scatter in the observed and
simulated GCVFs.  The observations display a significantly larger
scatter than the simulations.  Uncertainties in the measurement of
$V_{{\rm cl}}$ and Poisson errors both contribute to the observed
scatter.  Taking the plot of $\log_{10}[N(>(V_{{\rm c}}/V_{{\rm
cl}}))]$ versus $V_{{\rm c}}/V_{{\rm cl}}$ to be a power law, the
variance in $\log_{10}[N(>(V_{{\rm c}}/V_{{\rm cl}}))]$ due to measurement errors in $\sigma_{{\rm cl}}$
is

\begin{equation}\Delta_{\log_{10}[N(>(V_{{\rm c}}/V_{{\rm cl}}))]}^2 =
\frac{1}{\ln 10} \frac{\Delta_{{\rm {V_{cl}}}}^2}{V_{{\rm cl}}^2}, 
\end{equation}

\noindent where $V_{{\rm cl}} = {\sqrt 2}\sigma_{{\rm cl}}$.  Using
the values of $\sigma_{{\rm cl}}$ found in \S{\ref{sec:global}}, and
computing $\Delta_{\rm {V_{cl}}}$ from the errors in $\sigma_{{\rm
cl}}$, we found that the scatter due to measurement errors in
$\sigma_{{\rm cl}}$ is 0.009, while the scatter due to Poisson noise
is 0.012.  The total observed scatter is 0.064.  Thus, the instrinsic
scatter is $\sim$0.043.  This is very similar to the intrinsic scatter
we measured in the simulations.  Thus our observed GCVFs show the same
degree of variations found in $\Lambda$CDM dark matter cluster
simulations.

\subsection{Trends in the Cluster GCVF with Cluster Velocity Dispersion}
\label{sec:trends}

\begin{figure}
\psfig{figure=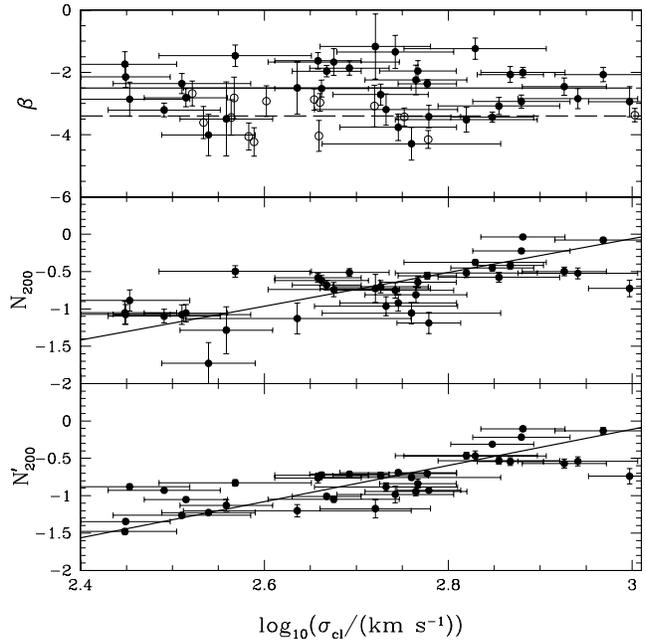,width=3.5truein,height=3.5truein}
\caption{Each observed and simulated cluster GCVF was fit to the
  equation $\log_{10} dN/dV_{{\rm c}} = \log_{10}N_{200} +
  \beta\log_{10}\left( V_{{\rm c}} / 200 {\rm ~km s^{-1}}\right)$.
  {\em Top Panel:} The slope of the cluster GCVF as a function of
  cluster velocity dispersion.  Solid points are for the observed
  cluster GCVFs.  The mean value is indicated by the solid line.  Open
  points are for the simulated cluster GCVFs.  The mean value is
  indicated by a dashed line.  {\em Middle Panel:} The normalization
  of the observed cluster GCVF as a function of cluster velocity
  dispersion.  High $\sigma_{{\rm cl}}$ clusters tend to contain more
  galaxies than low $\sigma_{{\rm cl}}$.  The line that best fits this
  trend is given by Equation \ref{equation:normtrend} and is
  overplotted. {\em Bottom Panel:} The normalizations of the fits for
  the observed cluster GCVFs were recomputed while holding the slope
  fixed to the mean value shown in the top panel.  The new
  normalizations are plotted against $\sigma_{{\rm cl}}$ in the bottom
  panel.  The line that best fits these data is given by Equation
  \ref{equation:newnormtrend} and is overplotted.}
\label{sigmatrends} 
\end{figure}

Figure \ref{sigmabins} shows that the cluster GCVF depends upon the cluster velocity dispersion.
To quantify the observed cluster GCVF and
its dependence on $\sigma_{\rm{cl}}$, we fitted each with a function of
the following form:

\begin{equation}
\log \left[\frac{dN(V_{{\rm c}})}{dV_{{\rm c}}}\right] = \log_{10}N_{200} +
\beta\log_{10}\left(\frac{V_{{\rm c}}}{200 {\rm ~km s^{-1}}}\right),
\label{clusterfit}
\end{equation}

\noindent where $\beta$ represents the slope of the power law and
$N_{{\rm 200}}$ is the normalization at $V_{{\rm c}}$=200$\kms$.  The
results of the fit are reported in Table \ref{datatable}, columns
8--11, and are plotted against $\log_{10}\sigma_{cl}$ in Figure
\ref{sigmatrends} (filled points).  The top panel of Figure
\ref{sigmatrends} shows that the slope, $\beta$, while showing a fair
amount of scatter, does not vary systematically with $\sigma_{{\rm
cl}}$.  It has a mean value of $\overline{\beta} = -2.5\pm0.8$.  The
slope of the composite cluster GCVF shown in Figure \ref{composite} is
$-2.61\pm0.04$, consistent with the mean slope.  The normalizations of
the individual cluster GCVFs are plotted in the middle panel of Figure
\ref{sigmatrends} and show a clear systematic trend in the sense that
higher mass clusters host more galaxies.  The best fit line to these
data are

\begin{equation}
\log_{10}N_{200}=(2.26\pm0.2)\log_{10}\sigma_{{\rm cl}} - (6.84\pm0.6).
\label{equation:normtrend}
\end{equation}

\noindent  Finally, we recalculated our normalizations with the slope fixed to
its mean value of $-2.5$.  These values of $N_{200}^\prime$ can be found in Table
\ref{datatable}, columns 12 and 13, and are plotted against
$\sigma_{{\rm cl}}$ in the third panel of Figure \ref{sigmatrends}.  The best
fit line to these data are

\begin{equation}
\log_{10}N_{200}^\prime = \left(2.42\pm0.2\right) \sigma_{{\rm cl}} - \left(7.37\pm0.5\right).
\label{equation:newnormtrend}
\end{equation}

\noindent This slope is somewhat smaller than theoretically expected,
but not at a statistically significant level.

We compared the slopes of the observed GCVFs to those of the simulated GCVFs.
The latter are plotted as open circles in the top panel of 
Figure \ref{sigmatrends}.  The dashed
line is drawn at their mean value: $\overline{\beta_{sim}} =
-3.4\pm0.8$.  The average value of the simulations is somewhat steeper
than that of the observations, although formally they agree within the
uncertainties.  It should be noted that the slopes are not directly
comparable because the simulations do not account for the effects of
baryons.  The effects of baryons on the slope of
the GCVF are discussed in the next subsection.

\subsection{The Effects of Baryons on the Cluster GCVF}
\label{sec:hires}

We found in \S{\ref{sec:trends}} that subhalos in simulated clusters
have steeper GCVFs than measured for galaxies in clusters.  We also
noted that a more meaningful comparison between observations and
simulations would require a determination of the circular velocity
profile taking into account the effects of both dark matter {\em and}
baryons.  Baryons affect the circular velocity profile of a galaxy in
two ways.  First, they significantly contribute to the enclosed mass at
small radii.  Second, they adiabatically pull the dark matter into a
more concentrated equilibrium configuration than it would have in
their absence.  The total circular velocity profile is therefore given
by

\begin{equation}
V_{{\rm c}}^2(r) = V_{{\rm c,b}}^2(r) + V_{{\rm c,DM}}^2(r),
\end{equation}

\noindent where $V_{{\rm c,b}}(r)$ is the circular velocity profile of
the collapsed baryonic mass and $V_{{\rm c,DM}}(r)$ is the circular
velocity profile of the adiabatically contracted dark matter halo.

The degree of these effects depends on the mass distribution of the
baryons.  The challenge lies in populating the subhalos with galaxies
in a way that mimics, as closely as possible, observed empirical
relations.  A full description of the baryonic mass distribution
requires knowledge not only the density profile shape (galaxy type),
but also a normalization (total galaxy mass) and a scalelength (galaxy
size).  Our method for populating the subhalos identified in our
highest resolution simulation is described in Appendix
\ref{sec:appendix}.  Given the density profiles of the baryonic
component of each subhalo, their circular velocity profiles
were computed.  Hernquist profiles are spherical, so the rotation
curves of early type galaxies were straightforward to calculate:
$V_{{\rm sph}} = GM(r)/r$.  Determining the rotation curve of
exponential profiles was also staightforward, but required taking
account of the non-spherical geometry.

The determination of $V_{{\rm c,DM}}(r)$ required modelling how the
dark matter responds to the collapse of the baryons.  We assumed that
the baryons contract slowly and that the halo remains spherical.
Under these conditions, a dark matter particle conserves angular
momentum as it moves closer to the center of the potential
\citep{Blumenthal86, Dalcanton97, Mo98}:

\begin{equation}
M_{{\rm f}}(r_{{\rm f}})r_{{\rm f}} = M_{{\rm i}}(r_{{\rm i}})r_{{\rm i}},
\label{equation:acontract}
\end{equation} 

\noindent where $r_{{\rm i}}$ is the initial mean radius of the
particle, $r_{{\rm f}}$ is the final mean radius of the particle, and 

\begin{equation}
M_{{\rm f}}(r_{{\rm f}}) = M_{{\rm b}}(r_{{\rm f}}) + M(r_{{\rm i}})(1 - f_{{\rm cool}}).
\label{equation:Mfinal}
\end{equation}

\noindent Equations \ref{equation:acontract} and \ref{equation:Mfinal}
both require knowledge of $M(r_{{\rm i}})$, the initial mass profile
of the dark matter halo, before adiabatic contraction.  While $V_{{\rm
pk}}$ was robustly measured from all of the simulations listed in Table
\ref{simtable}, a reliable determination of $V(r_{{\rm i}})$ or
$M(r_{{\rm i}})$ could only be made from the simulation described by the
last entry in Table \ref{simtable}, which has both very high
resolution and a small softening scale.  Using the measured $M(r_{{\rm
i}})$, the values of $f_{{\rm cool}}$ described in
\S{\ref{sec:assignmass}}, and the $M_{{\rm b}}(r)$ computed from the
density profiles described in Appendix \ref{sec:appendix}, we
calculated $M({{\rm r_f}})$ from Equations \ref{equation:acontract} and
\ref{equation:Mfinal}.  From $M({{\rm r_f}})$ it was trivial to
calculate $V_{{\rm c,DM}}(r)$.

Once the total circular velocity profile $V_{{\rm c}}(r)$ due to dark
matter and baryons was determined, we measured a
characteristic circular velocity, $V_{{\rm c}}$, that is analogous to
that deduced for our observed early and late type galaxies.  We
estimated $V_{{\rm c}}$ as the circular velocity at three scale
lengths.

\begin{figure}
\psfig{figure=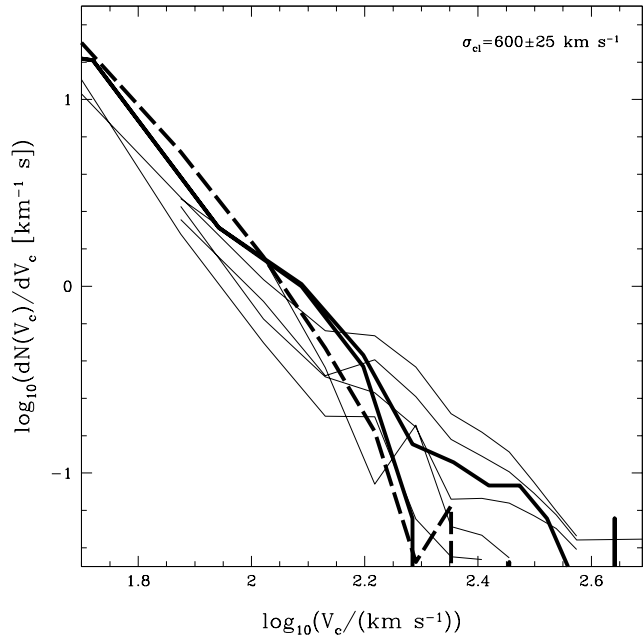,width=3.5truein,height=3.5truein}
\caption{Comparison of the GCVF measured from our highest resolution
  galaxy cluster simulation (HR Virgo) to that measured in SDSS clusters.  The
  dashed line is the simulated GCVF constructed using the $V_{{\rm pk}}$ values
  output by {\sc SKID}.  Thick solid lines represent the range of
  corrected GCVFs given uncertainties in how gas infall affects
  galaxy rotation curves (see Appendix \ref{sec:appendix} for details).
  Thin solid lines are the GCVFs of SDSS clusters with measured
  velocity dispersions within 25 $\kms$ of that of the simulated
  cluster.
  \label{simsvobs}}
\end{figure}

In Figure \ref{simsvobs}, we compare the simulated GCVF to that of
SDSS clusters that have similar velocity dispersions.  The dashed line
represents the simulated cluster GCVF in the limit of no baryons and assuming that $V_{{\rm c}} =
V_{{\rm pk}}$.  The simulated cluster GCVF in this case is too steep,
underpredicting the number of galaxies at high $V_{{\rm c}}$.  The
bold solid lines bracket simulated GCVFs which have been corrected for
the effects of baryonic infall.  We varied the total mass that
infalls to form ellipticals ($0.025 < f_{{\rm cool}} < 0.1$) and the
peak circular velocity above which subhalos host early type rather
than late type galaxies ($125\kms < V_{{\rm break}} < 150\kms$).  See
Appendix {\ref{sec:appendix}}.  When the effects of baryons were included, the
number of high $V_{{\rm c}}$ galaxies increased, while the number of
low $V_{{\rm c}}$ galaxies decreased, bringing the simulations into
rough agreement with the observations.  This effect held whether
$V_{{\rm c}}$ was measured at 2, 3, or 4 scale lengths, but was
strongest if $V_{{\rm c}}$ was measured at 2 scale lengths.

While $\Lambda$CDM models can reproduce the cluster GCVF, they
overproduce the number of field galaxies with $V_{{\rm }} \la 120
\kms$ \citep{Gonzalez00, Kochanek01}.  Any solution to the
disagreement in the field must preserve the agreement found in
clusters.  Modifications to CDM would affect the substructure within
clusters as well as in the field, and tend to contradict other
observations \citep{Barkana01, Gnedin01, Hui01, Miralda02}.  Supernova
feedback alone fails to reproduce the luminosity function of the Local
Group \citep{Somerville02}, but a combination of feedback and
squelching can.  However, squelching cannot play a role at the
circular velocities we have investigated here.  It affects only
galaxies with circular velocities $\la$ 50 $\kms$.  The importance of
supernova feedback at higher circular velocities is uncertain.  Recent
work by \citet{MacLow99} suggests that it can only regulate star
formation, not remove mass at these scales.  Regardless, it should be
equally important in clusters and in the field.  Finally, any
incompleteness that exists in our cluster GCVF should also exist in
our field GCVF.

\subsection{The Cluster Versus Field GCVF}
\label{sec:field}

The circular velocity function of dark matter halos within clusters is
expected to be a function of environment.  Although we found only
small differences among the populations of massive clusters, there may
still be substantial differences between clsuters and the field.
First, hierarchical clustering theory predicts that massive halos form
from rarer density peaks and should therefore be more clustered than
low mass halos \citep{Kaiser84, Davis85, White87, White88,
Governato98}.  Second, halos within clusters are subject to the
mass-altering processes of tidal stripping, dynamical friction, and
galaxy harrassment \citep{Moore96}.  Third, low mass halos in dense
regions formed, on average, earlier than those in the field.
Reionization can therfore preferentially suppress the late-time
formation of galaxies within low mass halos in the field
\citep{Quinn96, Tully02, Benson02, Somerville02}, although at
velocites below what we can probe in this work.

\citet{Cole89} determined the circular velcoity function of field galaxies
by starting with a Schechter luminosity function and assuming $L
\propto V_c^{3-4}$, not taking into account that a late type galaxy
has a lower circular velocity than an early type with the same
absolute magnitude, or that their relative abundances depend upon luminosity.  \citet{Shimasaku93}
computed the field GCVF galaxy-by-galaxy using both HI 21 cm line
widths ($\sim$400) and the Tully-Fisher relation ($\sim$300) for
spiral galaxies and stellar velocity dispersions ($\sim$100) and the
Faber Jackson relation ($\sim$90) for early type galaxies.  They found
that early types dominate at the high velocity end.

\citet{Gonzalez00} determined the field GCVF by performing variable
transformations on the type-specific B-band luminosity functions
determined from the SSRS2 survey.  They found that, given the
available Tully-Fisher and Faber Jackson Relations, all field galaxies
could be treated as late types without changing the results.  Doing
so, they found that the number density of galaxies per unit velocity
can be described by a Schecter function,

\begin{equation}
\tilde{\Psi}(V_{{\rm c}})dV_{{\rm c}} = \tilde{\Psi}_{\ast} \left(\frac{V_{{\rm c}}}{V_{{\rm c},\ast}}\right)^{\beta} \exp\left[
  -\left(\frac{V_{{\rm c}}}{V_{{\rm c},\ast}}\right)^{n}\right] \frac{dV_{{\rm c}}}{V_{{\rm c},\ast}},
\label{field}
\end{equation}

\noindent where $\tilde{\Psi}_{\ast} = (3.2\pm0.6)\times10^{-2}$
Mpc$^{-3}$ $h^3$, $\beta = -1.3\pm0.13$, $n = 2.5$, $V_{{\rm c},\ast} =
247\pm7$.  \citet{Kochanek01} found similar
results, although they took into account the different transformations
for early and late types.  \citet{Sheth03} analyzed field early type
galaxies within the SDSS, and found that accounting for the scatter
around the mean Tully-Fisher and Faber Jackson or Fundamental Plane
relations is extremely important for reproducing the high velocity end
of the GCVF.

\begin{figure}
\psfig{figure=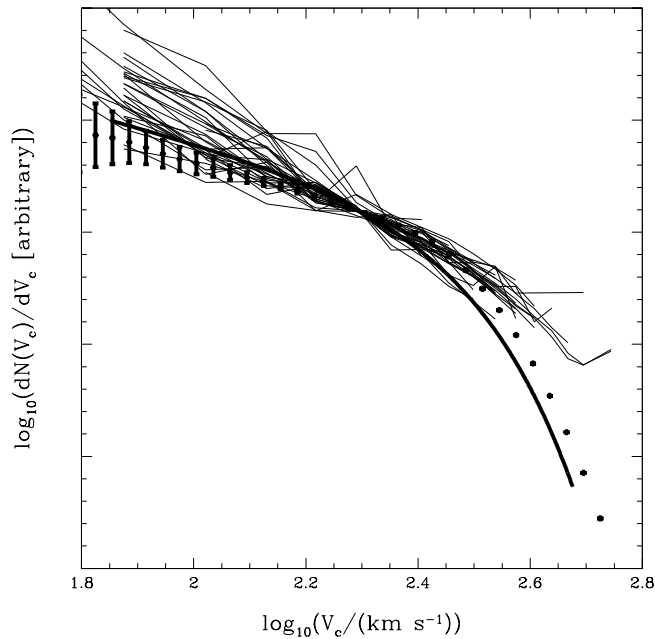,width=3.5truein,height=3.5truein}
\caption{GCVFs for the SDSS cluster sample and the field.  The thin
lines represent the GCVFs for SDSS clusters, normalized at
$\log_{10}(V_{{\rm c}}/\kms) \approx 2.3$.  The solid bold line is the
similarly normalized observed field GCVF as determined by
\citet{Gonzalez00}.  The points with error bars are the field GCVF
determined from SDSS data as described in \S{\ref{sec:field}}.
Because of the arbitrary normalization, this plot is suitable only for
assessing differences in the shape of the GCVF in clusters and the
field.  Although the cluster GCVFs display a range of slopes, they are
steeper than the field for $V_{{\rm c}} \la 200$ and more shallow for
$V_{{\rm c}} \ga 200$.  The field GCVFs are consistent with one
another within the error bars at $V_{{\rm c}} < 200 \kms$.  The field
GCVF we determined lies above that measured by \protect\citet{Gonzalez00} at
$V_{{\rm c}} > 200 \kms$ because we took into account the scatter
around the relations used to compute $V_{{\rm c}}$ for early type
galaxies, which dominate the GCVF at large $V_{{\rm c}}$.}
\label{plotallclusters} 
\end{figure}

In order to allow a direct comparison to our cluster GCVFs, we
constructed a field GCVF using the same relations that we used in the
previous cluster analysis.  We select galaxies from the SDSS EDR
spectroscopic sample with 14.5 $>$ $r^*$ $>$ 17.77.  Each of the
resulting $\sim$29,000 galaxies was assigned a weight equal to the
inverse of the volume over which the galaxy could be observed, given
the apparent magnitude limits.  Circular velocities were computed for
each galaxy as described in \S{\ref{sec:late}} and \S{\ref{sec:early}}.  The scatter was dealt
with as described in \S{\ref{sec:scatter}}.  The value of the field
GCVF in a given circular velocity bin is equal to the sum of the
weights of the galaxies in that bin.  The results are plotted as
points with error bars in Figure \ref{plotallclusters}, along with the
functional forms for the field GCVF given by \citet{Gonzalez00} (Equation \ref{field}; solid
bold line).  Figure \ref{plotallclusters} illustrates that the cluster GCVF
shows much less curvature than that of the field, being steeper for
$V_{{\rm c}} \la 200 \kms$ and more shallow for $V_{{\rm c}} \ga
200 \kms$.


Following the discussion in Section \ref{sec:trends}, the difference
in the shapes of the field and cluster GCVFs can be explained by their
different relative abundances of early and late type galaxies as a
function of luminosity.  It must be noted that our division of
galaxies into early and late types depends upon the $u^* - r^*$ color,
which is sensitive to star formation.  There is some evidence that the
cluster environment supresses star formation \citep{Couch87, Barger96,
Poggianti99}.  A field galaxy that falls into a cluster and
experiences truncated star formation would redden, but its $r^*$-band
luminosity would not change significantly, because $r^*$ is sensitive
to the total stellar mass.  If the change in $u^* - r^*$ is large
enough to change late types into early types, the cluster GCVF would
have more high-velocity galaxies and fewer lower-velocity galaxies
relative to the field, as is observed.  While the trend is correct,
detailed models are required to determine to what extent this effect
can explain the observed differences between the field and cluster
GCVFs.

\section{Summary}
\label{sec:summary}

We constructed the cluster and field galaxy circular velocity
functions using data from the Sloan Digital Sky Survey Early Data
Release.  Circular velocities were determined using the inverse Tully-Fisher and Fundamental Plane relations on a galaxy-by-galaxy basis,
rather than by performing a transformation of variables on a
functional fit to the luminosity function.  For comparison to the
observations, we also analyzed the GCVFs determined from 15 clusters
simulated in the $\Lambda$CDM concordance cosmology.  Our results can
be summarized as follows:

1.  We found that, although the observed cluster luminosity function is a Schechter function,  the observed cluster GCVF can be described as a
power law.

2.  We tested the self-similarity claimed by \citet{Moore99}
for our scaled simulated and observed cluster GCVFs.  The simulated
GCVFs are indeed very similar to one another, and show only a small
scatter.  The scatter in the observed cluster GCVFs is much
larger, but can be explained by measurement error.  Thus, the
intrinsic scatter in the observed cluster GCVF is similar to that
predicted.

3.  The normalization of the observed cluster GCVF increases with
cluster velocity dispersion, while the slope is constant.  The average
slope of the observed cluster GCVF is somewhat flatter ($\beta \approx
-2.5$) than that computed from $\Lambda$CDM cluster simulations
($\beta = -3.4$).  However, the errors on these average slopes are
large, and overlap.  If the difference between the simulated and the
observed clusters is real, it is most likely due to the fact that the
simulations do not include the effects of baryons on galaxy rotation
curves.

4.  We estimated the effects of baryons on the simulated GCVFs.
Baryons can flatten the simulated GCVF enough to bring it into
agreement with the GCVFs of observed clusters with similar velocity
dispersions.  This agreement between the amount of cluster
substructure predicted by N--body simulations and that observed in
real clusters has been found previously by \citet{Moore99} for an SCDM
simulation of a Virgo-like cluster.  Our approach is different in that
we treat early and late types distinctly, as well as account for the
effects of baryons on galaxy rotation curves.  In addition, our
simulated cluster has evolved in the concordance cosmology.


\section*{acknowledgments}

VD and DR acknowledge funding from  the Graduate Student Researchers Program.

JJD was partially supported through the Alfred P.\ Sloan Foundation.

TRQ was partially supported by the National Science Foundation.

Simulations  were run  at the  ARSC (Fairbanks)  and  CINECA (Bologna,
Italy) supercomputing centers.

This  research has made  use of  the NASA/IPAC  Extragalactic Database
(NED), which  is operated by  the Jet Propulsion  Laboratory, Caltech,
under contract with the National Aeronautics and Space Administration.

The  Sloan  Digital  Sky Survey  (SDSS)  is  a  joint project  of  The
University of Chicago, Fermilab, the Institute for Advanced Study, the
Japan  Participation  Group, The  Johns  Hopkins  University, the  Los
Alamos  National Laboratory,  the  Max-Planck-Institute for  Astronomy
(MPIA),  the Max-Planck-Institute for  Astrophysics (MPA),  New Mexico
State University, the United  States Naval Observatory, the University
of Pittsburgh, Princeton University, and the University of Washington.
Apache Point Observatory, site of  the SDSS telescopes, is operated by
the Astrophysical Research Consortium (ARC).

Funding  for the  project has  been provided  by the  Alfred  P. Sloan
Foundation, the SDSS member institutions, the National Aeronautics and
Space   Administration,   the   National   Science   Foundation,   the
U.S. Department  of Energy, the  Japanese Monbukagakusho, and  the Max
Planck Society.  The SDSS Web site is {\tt http://www.sdss.org/}.



\appendix

\section{Populating Subhalos with Galaxies}
\label{sec:appendix}

In order to estimate the effects of baryons on galaxy circular
velocity profiles, we must know the circular velocity profiles of both
the dark matter halo and the embedded collapsed baryons.  A full
description of the baryonic mass distribution requires knowledge not
only of the density profile shape (galaxy type), but also a normalization
(total galaxy mass) and a scalelength (galaxy size).  In the remainder
of this section, we outline our method for assigning galaxy types,
masses, and sizes to the population of subhalos identified in the
zero-redshift output of our highest resolution cluster simulation.
Once the dark matter and baryon density profiles are fully specified,
it is straightforward to compute the final rotation curve due to both
components.

\subsubsection{Assigning Galaxy Types}
\label{sec:assigntype}

Although galaxies exhibit a range of bulge-to-disk ratios, for
simplicity we assign either a pure spheroidal or a pure disk baryonic
component to each subhalo in the HR Virgo simulation.  Spheroidals are
described by a Hernquist density profile \citep{Hernquist90}:

\begin{equation}
\rho(r) = \frac{M_{{\rm sph}}}{2\pi} 
\frac{R_{{\rm sph}}}{r} 
\frac{1}{(r+R_{{\rm sph}})^3},
\label{equation:hernquist}
\end{equation}

\noindent where $M_{{\rm sph}}$ is the total mass of the spheroid, and the
scalelength $R_{\rm sph}$ is related to the projected half-light radius $R_{\rm 0}$
of a deVaucoleurs surface brightness profile by $R_{\rm sph}$ = 0.551$R_{\rm 0}$.

Late type galaxies are assumed to be infinitely thin, and have surface density
profiles of the form

\begin{equation}
\Sigma(r) = \frac{M_{{\rm d}}}{2\pi R_{{\rm d}}^2} e^{-r/R_{{\rm d}}},
\label{equation:exponential}
\end{equation}

\noindent where $M_{\rm d}$ is the total mass of the disk and $R_{\rm
d}$ is the disk scale length.

\begin{equation}
M_{{\rm d}} = 2\pi \Sigma_{{\rm 0}} R_{{\rm d}}^2.
\end{equation}

Observationally, galaxy type is a function of circular velocity, with
early type galaxies dominating at high $V_{{\rm c}}$
\citep{Shimasaku93, Kochanek01, Sheth03}.  Only the peak circular
velocity of the dark matter component, $V_{{\rm pk}}$, can be measured
for each subhalo.  We therefore assume that if early type galaxies
dominate at high $V_{{\rm c}}$, they also dominate at high $V_{{\rm
pk}}$.  We randomly assign each subhalo the designation of early or
late type based upon the early type fraction as a function of $V_{{\rm
pk}}$, $f_{{\rm early}}(V_{{\rm pk}})$.  For simplicity we adopt a
step function for $f_{{\rm early}}(V_{{\rm pk}})$.

\begin{equation}
f_{{\rm early}}\left(V_{{\rm pk}}\right) = \left\{ \begin{array}
  {r@{\quad:\quad}l} 0 & V_{{\rm pk}}< V_{{\rm break}} \\ 1 & V_{{\rm pk}}\geq V_{{\rm break}} \end{array} \right.,  
\end{equation}

\noindent where $125\kms < V_{{\rm break}} < 150\kms$.  Alternative functional
forms which provide smoother transitions in the run of galaxy type
with $V_{{\rm pk}}$ do not produce significantly different results.
Subtantially different values of $V_{{\rm break}}$ produce
disagreement with the joint distribution of $R_{{\rm 0}}$ and
$\sigma_{{\rm 0}}$ found by \citet{Bernc03} (see
\${\ref{sec:earlysize}}).

\subsubsection{Assigning Galaxy Masses}
\label{sec:assignmass}

In the previous subsection, we described our procedure for assigning
either a Hernquist or exponential density profile to each subhalo.
Each of these two profiles is parametrized by both a mass in collapsed
baryons ($M_{{\rm sph}}$ or $M_{{\rm d}}$) and a scale length ($R_{\rm
sph}$ or $R_{\rm d}$).  We set the baryonic mass embedded within each
subhalo equal to a fraction of the virial mass of that subhalo at the
time that the baryons collapsed.

The first step towards determining the mass in cooled baryons embedded
within each subhalo is to determine the virial mass of each subhalo.
The virial radius of a simulated dark matter halo is usually taken to
be the $R_{\rm 200}$, the radius interior to which the average density
is equal to 200 times the critical density.  The virial mass is then
estimated as $M_{\rm 200}$, the mass interior to $R_{\rm 200}$.  For
subhalos, $R_{200}$, and therefore $M_{\rm 200}$ is very difficult to
measure.  In addition, the baryons likely collapsed prior to the
subhalo falling into the cluster and suffering the dynamical effects
therein.  We are therefore interested in the value of $M_{200}$ for
each subhalo before cluster incorporation.  Although the cluster
environment can lead to significant mass stripping, the peak circular
velocities of halos ($V_{\rm pk}$) remain relatively constant
\citep{Hayashi02}.  We can therefore use the values of $V_{\rm pk}$
measured in the redshift zero simulation output to infer $M_{200}$
using a correlation between the two found by \citet{NFW97}:

\begin{equation}
\log\left(\frac{M_{{\rm 200}}}{10^{10}M_{\sun}}\right) =
3.23\log\left(\frac{V_{{\rm pk}}}{\kms}\right) - 5.31.
\label{equation:MV}
\end{equation}

Once $M_{{\rm 200}}$ has been determined for each subhalo, the second
step is to specify the fraction, $f_{\rm cool}$ of $M_{200}$ that
winds up as cooled baryons.  For late-type galaxies, we adopt the
fitting function found by \citet{Gonzalez00} in an analysis of the
semi-analytic models of \citet{Somerville99}:

\begin{equation}
f_{{\rm cool}} \simeq 0.1 \left(\frac{x-0.25}{1+x^2}\right),
\end{equation}

\noindent where $x = V_{{\rm pk}} / (200 \kms)$.  The above is valid
for late type galaxies with $V_{{\rm pk}} \simeq$ 60 -- 350 $\kms$.
For subhalos with $V_{{\rm pk}} \le 60 \kms$, we set $f_{{\rm cool}} =
0.02$.  This minimum value of $f_{{\rm cool}}$ does not affect our
comparison to observations, as subhalos with $V_{{\rm pk}} \le 60
\kms$ are unlikely to have $V_{{\rm c}} > V_{{\rm compl}}$.  For early
type galaxies, we explore $f_{{\rm cool}} = 0.025, 0.05, 0.1$.


\subsubsection{Early Type Galaxy Sizes}
\label{sec:earlysize}

In the previous two sections, we presented simple prescriptions for
assigning a density profile shape and normalization to each subhalo
within our highest resolution cluster simulation.  Before the effects
of baryons on galaxy rotation curves can be explored, we must assign
scale lengths to the baryonic component of each subhalo.  One
projection of the Fundamental Plane tells us that the line of sight
velocity dispersions, $\sigma_{\rm los}$, or early type galaxies are
correlated with their half-light radii, $R_e$, which are in turn
simply related to the scale length of a Hernquist density profile.
Under the assumption that $\sigma_{\rm
los}$ is also correlated with the peak circular velocities $V_{\rm
pk}$ of the dark matter halos in which they form, we can relate
$R_{\rm sph}$ to $V_{\rm pk}$.

In order to assign values of $R_{\rm sph}$ to each subhalo, we first
assign values of $R_{\rm 0}$.  Using the full covariance matrix from
B03, we draw one ($\sigma_{{\rm los}}$, $R_{{\rm 0}}$) pair for each
subhalo selected through the procedure described in
\S{\ref{sec:assigntype}} to host an early type galaxy.  Each value of
$R_{{\rm 0}}$ is assigned a rank, with the largest $R_{{\rm 0}}$
receiving a rank of 1.  Ranks are similarly assigned for each drawn
$\sigma_{{\rm los}}$ and for each subhalo based opon its measured
$V_{{\rm pk}}$.  Although we do not know the relationship between
$\sigma_{{\rm los}}$ and $V_{{\rm pk}}$, we expect both to correlate
with $R_{{\rm 0}}$.  Thus, we assign each ($\sigma_{{\rm los}}$,
$R_{{\rm 0}}$) pair to a subhalo that has the same rank as that of the
$\sigma_{{\rm los}}$ in that pair.  This method of assigning sizes to
subhalos preserves the scatter in the correlation between $\sigma_{\rm
los}$ and $R_{{\rm 0}}$, but assumes a perfect correlation between
$\sigma_{\rm los}$ and $V_{\rm pk}$.

Once values of $R_{\rm 0}$ have been assigned to each subhalo, $R_{\rm
sph}$ is determined through $R_{\rm sph} = 0.551R_0$.
 
\subsubsection{Late Type Galaxy Sizes}

Our statistical assignment of early type galaxy sizes to subhalos was driven by observed correlations.  For disk galaxies,  
if baryons initially share the spatial and angular momentum
distribution of their host dark matter halo, and conserve angular
momentum as they collapse, the
resulting scalelength is related to the properties of the dark matter
halo \citep{Dalcanton97, Mo98}.  We use this idea to assign a scale
length, $R_{{\rm d}}$, to each subhalo chosen to host a late type
galaxies.  \citet{Mo98} show that

\begin{equation}
R_{\rm d} = \frac{1}{\sqrt{2}} \frac{j_{\rm d}}{f_{{\rm cool}}} \lambda r_{{\rm 200}} f_c^{-1/2} f_R, 
\end{equation}

\noindent where $j_d$ is the fraction of the total halo angular momentum contained in the disk; $f_{\rm cool}$ is, as before, the fraction of the dark matter mass contained in cooling baryons; $\lambda$ is the spin parameter of the dark matter halo; and $r_{200}$ is the virial radius.  Additionally,  

\begin{equation} 
f_c \simeq \frac{2}{3} + \left(\frac{c}{21.5}\right)^{0.7}, 
\end{equation}

\noindent where $c$ is the halo concentration and 

\begin{eqnarray}
f_R \simeq \left(\frac{f_{{\rm cool}}}{j_{\rm d}}
\frac{\lambda}{0.1}\right)^{(-0.06 + 2.71f_{{\rm cool}} + 0.0047
  f_{{\rm cool}}/(j_{\rm d} \lambda))} \\ \times \left( 1 - 0.019c + 0.00025c^2 + 0.52/c\right)\end{eqnarray}

\noindent We have specified $f_{{\rm cool}}$ in \S{\ref{sec:assignmass}}, and we assume that $j_{\rm d} = f_{\rm cool}$.    
\citet{Bullock01} analyze a 60 $h^{-1}$ Mpc simulation carried out
with the adaptive refinement tree N--body code, and found that the
average halo spin parameter, ${\overline \lambda} = 0.04$,
independent of $V_{{\rm pk}}$.  They also find the following relation
for halo concentration:

\begin{equation}
c \approx 14 \sqrt{\frac{V_{{\rm pk}}}{200 \kms}}
\end{equation}

\noindent Finally, $r_{{\rm 200}}$ is the radius within which the
average halo density is 200 times the critical density.  It can be
calculated straightforwardly from $M_{200}$, the mass enclosed within
this radius:

\begin{equation}
r_{200} = \left(\frac{3 M_{200}}{800\pi \rho_{{\rm crit}}}\right)^{1/3},
\end{equation}

\noindent where $\rho_{\rm crit,0} \approx 136 {\rm M}_{\sun} {\rm kpc}^{-3}$ is the critical density of the universe and $M_{200}$ is determined from Equation \S{\ref{equation:MV}}.
  
\label{lastpage}

\end{document}